\documentclass[letter,twocolumn]{jpsj2} 
\usepackage{graphicx}
\usepackage{color}
\title
{Supersolid State of Ultracold Fermions in Optical Lattice}

\author
{
Akihisa Koga,$^1$ Takuji Higashiyama,$^2$ Kensuke Inaba,$^2$
Seiichiro Suga,$^2$ and Norio Kawakami$^{1,2}$
}

\inst{
$^1$Department of Physics, Kyoto University,
Kyoto 606-8502, Japan\\
$^2$Department of Applied Physics, Osaka University, 
Suita, Osaka 565-0871, Japan
}

\abst{
We study ultracold fermionic atoms trapped in an optical lattice 
with harmonic confinement by dynamical mean-field approximation. 
It is demonstrated that a supersolid state, 
where an $s$-wave superfluid coexists with a density-wave state 
with a checkerboard pattern, 
is stabilized by attractive onsite interactions on a square lattice. 
Our new finding here is that a confining potential plays an invaluable role
in stabilizing the supersolid state. 
We establish a rich phase diagram at low temperatures, 
which clearly shows how an insulator, a density wave and a superfluid 
compete with each other to produce an interesting domain structure. 
Our results shed light on the possibility of the supersolid state 
in fermionic optical lattice systems.
}

\kword{dynamical mean field theory, supersolid state}

\begin{document}
\maketitle


Ultracold atomic gases have 
attracted much interest~\cite{Review}
since the successful realization of Bose-Einstein condensation 
in a bosonic $\rm ^{87}Rb$ system~\cite{Rb}. 
Optical lattices, formed by loading ultracold atoms in a periodic potential, have been providing an ideal stage 
for experimental and theoretical studies of fundamental problems 
in condensed matter physics~\cite{BlochGreiner,Bloch,Jaksch,Morsch}.
Owing to its high controllability in interaction strength, 
particle number, and other parameters, 
many remarkable phenomena have been observed 
such as the phase transition between a Mott insulator and a superfluid 
in bosonic systems~\cite{Greiner}. 
More recently, a fermionic gas in the optical lattice has been a topic of 
extensive study, which has successfully lead to 
the observation of superfluidity 
in the case of attractive interactions~\cite{Chin}.

One of the interesting questions for such a fermionic optical lattice is 
how an $s$-wave superfluid (SSF) state coexists or competes 
with a density wave (DW) state, 
where the latter can be regarded as a sort of solid state. 
This provides an important issue in condensed matter physics, 
since it is directly related to a hot topic of current interest, 
the so-called supersolid state.  The existence of the supersolid state was
suggested for $\rm ^4He$ experimentally~\cite{Kim}, and this pioneering work
has stimulated theoretical investigations on bosonic systems~
\cite{SSboson1,SSboson2,SSboson3,SSboson4,SSboson5,SSboson6,SSboson7}
and Bose-Fermi mixtures~\cite{SSmix}. 
As for fermionic systems,
the optical lattice can be a potential candidate for it. 
However, it has not been clarified 
how the supersolid is stabilized in the optical lattice
except for the one-dimensional system\cite{Pour,Xianlong}
although the existence of the SSF and DW states 
has been discussed~\cite{FFLO,Machida,Dao,Burkov}. 

According to previous studies of the attractive Hubbard model 
on a periodic lattice without 
a confining potential~\cite{Shiba,Scalettar,Freericks,Capone,Garg,Micnas},
in systems on bipartite lattices, except for a one-dimensional case, 
the DW and SSF ground states are degenerate at half filling, which means that the supersolid state might be realizable in principle. However, the degenerate ground state is unstable against perturbations. For example, away from  half filling, the supersolid state immediately changes to a genuine SSF state, where a BCS-BEC crossover has been discussed~\cite{Garg,Keller,Toschi}. 
In the optical lattices, we have an additional confining potential, 
which makes the situation different from the homogeneous bulk system. 
This naturally motivates us to  address the question 
whether the supersolid state can be realized in the optical lattice.

In this study, we demonstrate that the supersolid state can indeed be realized 
in a fermionic optical lattice with attractive interactions. 
In particular, it is found that a confining potential plays an important role 
in stabilizing the supersolid state; 
it makes the supersolid state robust against perturbations 
in contrast to that in homogeneous systems. 
This suggests that the fermionic optical lattice can be a potential candidate 
for the realization of the supersolid state.

We consider ultracold fermionic atoms, which may be described by the Hubbard model with confinement as 
\begin{eqnarray}
H=-\sum_{\langle i,j \rangle \sigma} t_{ij} c_{i\sigma}^\dag c_{j\sigma}
-U\sum_i n_{i\uparrow}n_{i\downarrow} 
+ V_0 \sum_{i\sigma} R_i^2 n_{i\sigma},
\end{eqnarray}
where $c_{i\sigma} (c_{i\sigma}^\dag)$ annihilates (creates) a fermion 
at the $i$th site with spin $\sigma$ and 
$n_{i\sigma} = c_{i\sigma}^\dag c_{i\sigma}$.
$t_{ij}$ is the nearest-neighbor hopping, $U$ the attractive interaction, and
$V_0$ the curvature of a harmonic potential.  
Here, $R_i$ is the distance measured from the center of the system.

The ground-state properties of the Hubbard model 
on inhomogeneous lattices have theoretically been studied 
by various methods such as Bogoljubov-de Gennes equations~\cite{FFLO}, 
Gutzwiller approximation~\cite{Yamashita,Ruegg}, and
variational Monte Carlo simulations.\cite{Fujihara} 
Although ordered states are described properly in these approaches, 
it may be difficult to deal with the coexisting phase like a supersolid 
in the strong correlation regime. 
The density matrix renormalization group 
method~\cite{Machida,Xianlong} and 
quantum Monte Carlo method~\cite{QMC,Pour} 
are efficient for one-dimensional systems, but may be difficult 
to apply to higher-dimensional systems.
We here use dynamical mean-field approximation (DMFA)~
\cite{DMFT1,DMFT2,DMFT3,DMFT4}, 
which incorporates local particle correlations precisely, 
thus enabling us to obtain reliable results 
if spatially extended correlations are negligible. 
In fact, the method has successfully been applied to 
some inhomogeneous correlated systems such as the surface~\cite{Potthoff} or 
interface of Mott insulators~\cite{Okamoto} and 
repulsive fermionic atoms.\cite{Helmes,Snoek} 
An advantage of this method is to treat the SSF and DW states on an equal footing in the strong correlation regime.

In the framework of DMFA, the lattice Green's function is described in terms of the site-diagonal self-energy $\hat{\Sigma}_i\left(i\omega_n\right)$ as
\begin{eqnarray}
&&\left[\hat{G}_{lat}^{-1}\left(i\omega_n \right)\right]_{ij}\nonumber\\
&&= 
\delta_{ij}\left[ i\omega_n \hat{\sigma}_0 
+ \left(\mu-V_0 R_i^2 \right)\hat{\sigma}_z 
-\hat{\Sigma}_i\left(i\omega_n\right)\right] -t_{ij}\hat{\sigma}_z,
\label{eq:lat}\end{eqnarray}
where $\hat{\bf \sigma}_z$ is the $z$ component of the Pauli matrix, $\hat{\sigma}_0$ the identity matrix, $\mu$ the chemical potential, $\omega_n = (2n+1)\pi T$ the Matsubara frequency, and $T$ the temperature. 
A DMFA self-consistent loop of calculations is iterated 
under the condition that the site-diagonal component of 
the lattice Green's function is equal to the local Green's function 
obtained from the effective impurity model~\cite{DMFT1,DMFT2,DMFT3,DMFT4}.
When DMFA is applied to our inhomogeneous system, 
it is necessary to solve the effective impurity models $L$ times by iteration, 
where $L$ is the system size. 
For this purpose, we use a two-site approximation~\cite{2site,Higashiyama}. 
Although the effective bath is replaced by only one site in the method,
it has the advantage in taking into account 
both low- and high-energy properties reasonably well 
within restricted numerical resources~\cite{Higashiyama,Okamoto}.

In the following, we consider the square lattice with harmonic confinement as a simple model for the supersolid. We set $t$ as a unit of energy, and fix the curvature of the potential and the total number of atoms as $V_0=0.023$ and $N\sim 300 (N_\sigma \sim 150)$. In our system,
 the distribution of particles and order parameters is spatially modulated, which is optimized in the framework of DMFA. We thus calculate the density profile  $\langle n_{i\sigma} \rangle = 2T\sum_{n=0}{\rm Re}[
 G_{i\sigma}\left(i\omega_n\right)]+\frac{1}{2}$ and the distribution of
the pair potential
$\Delta_i = 2T\sum_{n=0}{\rm Re} [F_i\left(i\omega_n\right)]$, 
where $G_{i\sigma}(i\omega_n) [F_i(i\omega_n)]$ 
is the normal (anomalous) Green's function for the $i$th site.  Note that $\Delta_i$ represents the order parameter for the SSF state.


\begin{figure}[htb]
\vskip -2mm
\begin{center}
\includegraphics[width=8cm]{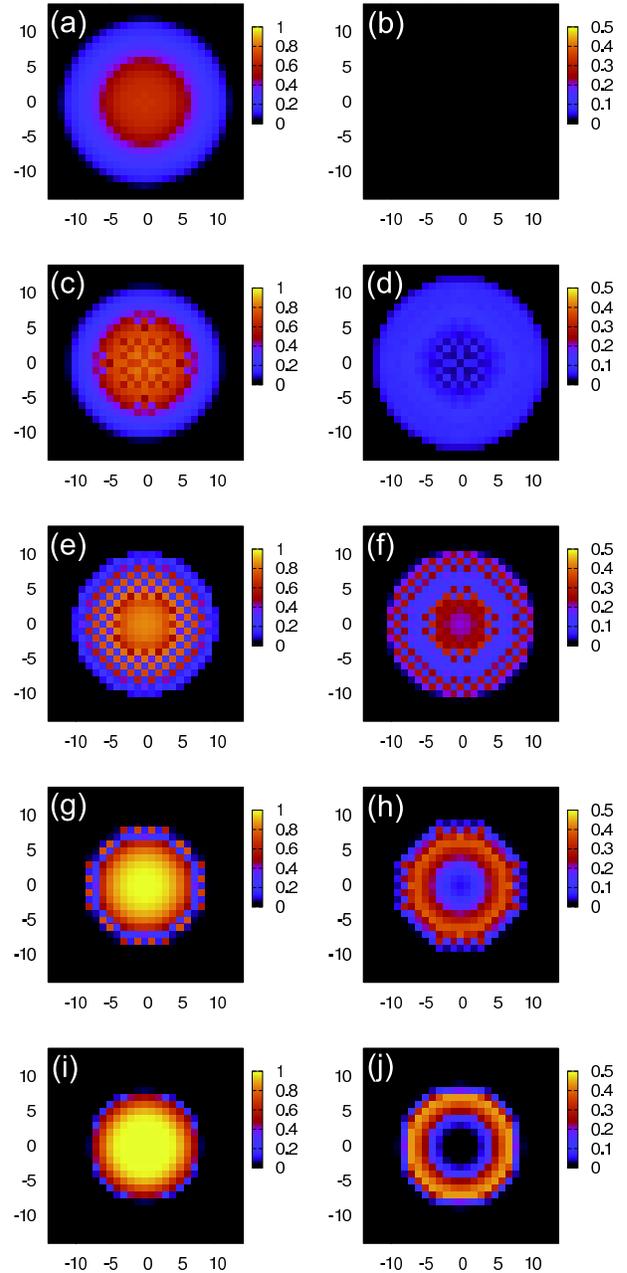}
\end{center}
\vskip -6mm
\caption{(Color online)
Density profile $\langle n_{i\sigma}\rangle$ (left panels) 
and pair potential $\Delta_i$ (right panels) on square lattice 
at $T=0.05$ when $U = 1.0, 3.0, 5.0, 7.0$ and $9.0$
(from top to bottom).
}
\label{fig:fig1}
\end{figure}

The obtained results at $T=0.05$ are shown in Fig. \ref{fig:fig1}. In the weak coupling case $(U=1)$, fermionic atoms are smoothly distributed up to $R\sim 13$,
where $R$ is the distance from the center of the harmonic potential. 
In this case, the pair potential is not yet developed 
[Figs. \ref{fig:fig1} (a) and (b)]. 
Therefore, a normal fluid state with short-range pair correlations emerges in the region $(R<13)$. Increasing the attractive interaction, fermions tend to gather around the bottom of the harmonic potential, as seen from $\langle n_{i\sigma} \rangle$ in Fig. \ref{fig:fig1} (c). Note here that the attractive interaction causes an SSF state with finite $\Delta_i$ in the region with $\langle n_{i\sigma} \rangle \neq 0$, as shown in Fig. \ref{fig:fig1} (d). 
This is consistent with the results obtained from 
the Bogoljubov-de Gennes equation~\cite{FFLO}. 
In this figure, we encounter a remarkable feature around the center of the harmonic potential $(R<5)$: a checkerboard structure appears in the density profile $\langle n_{i\sigma} \rangle$, while the SSF state is not suppressed completely even in the presence of the DW state. This implies that the DW state coexists with the SSF state, {\it i.e.},  
{\it a supersolid state is stabilized} in our optical lattice system. 
The characteristic properties of the supersolid state are clearly seen 
in the case of $U=5$, where the checkerboard structure appears 
in the doughnut-like region [Fig. \ref{fig:fig1} (e)].
Further increase in the interaction excludes the DW state out of the center.
It is seen in Figs. \ref{fig:fig1}(g) and \ref{fig:fig1}(i) that  fermionic atoms are concentrated around the bottom of the potential for large $U$.  In the region, two particles with opposite spins are strongly coupled by attractive interaction to form a hard-core boson, giving rise to an insulating state with $\langle n_{i\sigma}\rangle\sim 1$. 
We observe such behavior more clearly 
in Figs. \ref{fig:fig1}(h) and \ref{fig:fig1}(j). 

\begin{figure}[htb]
\vskip -2mm
\begin{center}
\includegraphics[width=7cm]{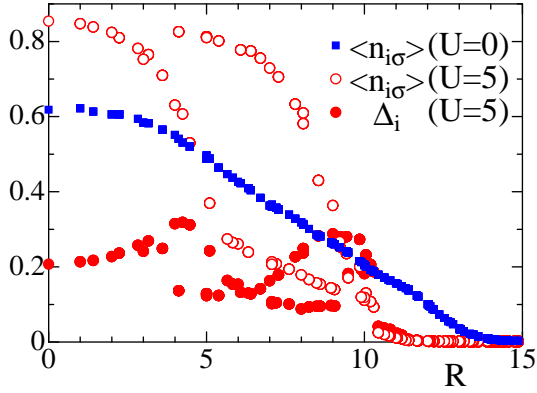}
\vskip -6mm
\end{center}
\caption{(Color online)
Open and solid circles represent $\langle n_{i\sigma}\rangle$ and $\Delta_i$ 
at $U=5$, and solid squares represent $\langle n_{i\sigma}\rangle$ 
in the noninteracting case at $T=0.05$.
}
\label{fig:R}
\end{figure}

To observe how the supersolid state is realized, we also show 
the spatial variations in $\langle n_{i\sigma} \rangle$ and $\Delta_i$ 
in Fig. \ref{fig:R} as functions of $R$. It is found that $\langle n_{i\sigma} \rangle$ and $\Delta_i$ describe smooth curves for $R<3$ and $10<R<11$, 
where the SSF state without the DW is realized. 
On the other hand, for $3<R<10$, two distinct amplitudes appear in $\langle n_{i\sigma} \rangle$, reflecting the fact that the DW state with two sublattices is realized. 
An important point is that the pair potential $\Delta_i$ is finite 
in the region 
although its profile is somewhat affected by the spatial variation in DW. 
We thus confirm that the supersolid state is realized in the doughnut-like region $(3<R<10)$.

\begin{figure}[htb]
\vskip -2mm
\begin{center}
\includegraphics[width=8cm]{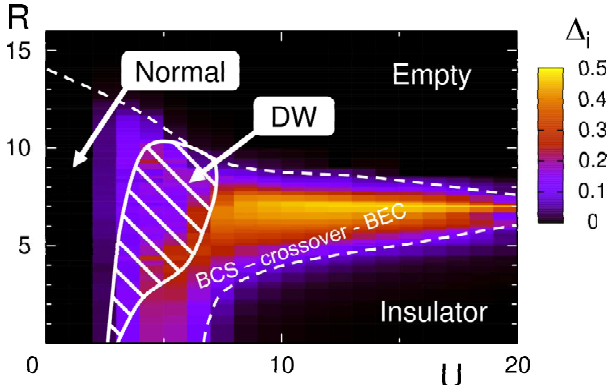}
\end{center}
\vskip -6mm
\caption{(Color online)
Phase diagram of attractive Hubbard model on optical lattice 
with $V_0=0.023$ and $N\sim 300$.
The density plot represents the amplitude of 
the $s$-wave pair potential $\Delta_i$. 
The DW state is realized in the shaded area. 
The broken lines serve as a visual guide which distinguishes the region 
with a fractional particle density from the empty 
and fully occupied regions.
}
\label{fig:phase}
\end{figure}

By performing similar calculations, we end up with the  phase diagram  shown in Fig. \ref{fig:phase}.
There are several remarkable features in the phase diagram. 
First, we notice that the supersolid state, characterized by the coexistence of SSF and DW orders, is indeed stabilized in a finite region ($U \simeq 3 \sim 7$), which is surrounded by the pure SSF state extended in a wider region. 
In the pure SSF region without the DW order, 
we still observe an interesting behavior, {\it i.e.}, a BCS-BEC crossover. 
When $U$ is rather small, the weak attractive interaction stabilizes a BCS-type SSF state, where $\Delta_i$ is induced in the whole region with $\langle n_{i\sigma} \rangle \neq 0$. 
In this parameter region, the pair potential is enhanced with 
an increase of the interaction $U$. 
On the other hand, in the strong coupling region, particles form 
short-range pairing states. 
In fact, most of the particles condense around the center 
yielding the insulating state, and the others form a BEC-type SSF state 
in the vicinity of $R=7$. 
Further increase in the attractive interaction narrows the SSF region, 
and suppresses the amplitude of the pair potential. 
Therefore, the pair potential $\Delta_i$ has a maximum at approximately 
$U=12$ and $R=7$,
 which may give a rough guide for the crossover region 
between the BCS-type and BEC-type SSF states.
We finally note that our supersolid state found in the weak coupling region 
at approximately $U=5$ is thus attributed to the coexisting state of 
the DW and BCS-type SSF states.

\begin{figure}[htb]
\vskip -2mm
\begin{center}
\includegraphics[width=8cm]{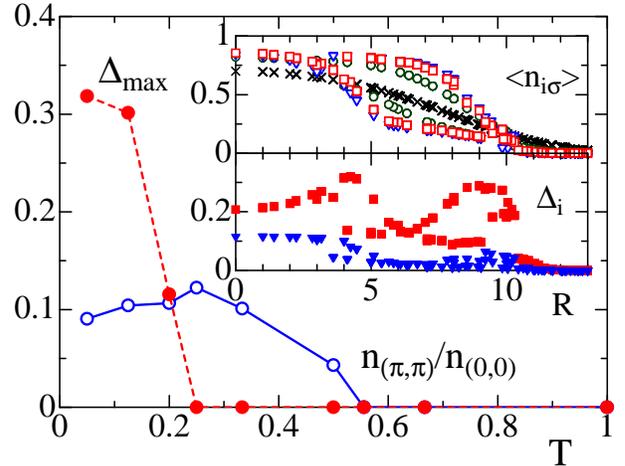}
\end{center}
\vskip -6mm
\caption{(Color online)
Maximum $\Delta_i$ and 
normalized $n_{(\pi,\pi)}/n_{(0,0)}$ 
when $U=5$ as functions of temperature $T$.
The inset shows $\langle n_{i\sigma} \rangle$ and $\Delta_i$ 
as functions of $R$.
Crosses, circles, triangles, squares represent the results at $T=1.0, 0.5, 0.2$ and $0.05$.
}
\label{fig:T}
\end{figure}

Next, we would like to discuss finite-temperature properties in more detail. 
Note that we employ mean-field approximation for ordered phases, so that the corresponding transition temperature is finite even in two dimensions. Nevertheless, some essential properties of the supersolid at finite temperatures can be captured by the preset treatment; the results may be applied to the case 
where a weak three dimensionality is introduced as should be in real experiments.
 Here, we focus on the case of $U=5$ to clarify how robust the supersolid state is against thermal fluctuations. The DW state is characterized by the checkerboard structure in the density profile $\langle n_{i\sigma} \rangle$, so that Fourier transform $n_{q}[=\sum \langle n_{i\sigma}\rangle \exp (iq R_i) ]$ at $q=(\pi,\pi)$ is an appropriate quantity to discuss its stability. In Fig. \ref{fig:T}, we show $n_{(\pi,\pi)}/n_{(0,0)}$ and the maximum of the pair potential $\Delta_{max}$.
Decreasing temperatures, $n_{(\pi,\pi)}/n_{(0,0)}$ becomes finite below $T_{DW}(\sim 0.5)$, where the DW state is realized. 
Once the DW is ordered, the corresponding spatial region with 
the checkerboard structure $(3<r<10)$ is hardly affected by temperature, 
as shown in the inset of Fig. \ref{fig:T}. 
On the other hand, the maximum pair potential $\Delta_{max}$ starts 
to develop at $T_{SSF}(\sim 0.2)$, which is lower than $T_{DW}$. 
Therefore, the supersolid state with finite $n_{(\pi,\pi)}$ and $\Delta_{max}$ is stabilized below $T_{SSF}$. 
These results imply that the DW state is more stable than the SSF state against thermal fluctuations for these parameters. On the other hand, around the border in the supersolid region in Fig. \ref{fig:phase}, the DW state becomes unstable. Therefore, to observe the supersolid state experimentally, it may be necessary to find appropriate parameters that should stabilize the DW state, since the SSF state is rather stable in the wide parameter region.


Finally, we emphasize the importance of the confining potential 
to realize the supersolid state in optical lattice systems. 
Let us recall that the DW state with a checkerboard pattern emerges 
when a commensurability condition, 
{\it e.g.}, half-filling, is satisfied for the particle density at least locally.
Therefore, in order to stabilize the supersolid state, it is essential to form a domain where the commensurability condition is met approximately. In our results presented here, such a domain is indeed formed in the doughnut-like region (Fig. \ref{fig:fig1}).
If the strength of the confining harmonic potential is decreased with $N/V\neq 1$ fixed, where $V$ is an effective system size, the transition temperature $T_{DW}$ approaches zero, and finally the supersolid state changes to a pure SSF state, since the domain that satisfies the half-filling condition disappears. 
Therefore, we claim that {\it a confining potential, 
which gives rise to an inhomogeneous distribution of the particle density, 
plays a key role in stabilizing a supersolid state} 
in the fermionic optical lattice. 
This in turn demonstrates that the optical lattice system could be a potential candidate for realizing a supersolid state.

We wish to comment on accessible experimental parameters 
for observing the supersolid state. 
The depth and curvature of the lattice and harmonic potentials 
can be controlled by adjusting the intensity and frequency of lasers. 
The hopping integral $t$ and the attractive interaction are then given as $t/E_R\sim 4\pi^{-1/2} (v_0/E_R)^{3/4} e^{-2(v_0/E_R)^{1/2}}$ and $U/E_R\sim -(8/\pi)^{1/2} a_s k_L (v_0/E_R)^{3/4}$, 
where $v_0$ and $k_L$ are the intensity 
and wave number of the laser for the lattice, $E_R$ the recoil energy, and $a_s$ the $s$-wave scattering length.\cite{Zwerger} In the paper, we have seen that the supersolid state appears in the vicinity of the BCS-BEC crossover region, which implies that the supersolid state is experimentally in accessible regions. Therefore, the supersolid state is expected to be observed by tuning these experimental parameters in the near future.

In summary, we have investigated the fermionic attractive Hubbard model 
in an optical lattice with harmonic confinement. 
Using DMFA, we have obtained a rich phase diagram on a square lattice, 
which has an interesting domain structure including the SSF state in the wide parameter region. In particular, we have found that the supersolid state, in which the SSF state coexists with the DW state, is stabilized at low temperatures.  It has also been elucidated that a confining potential plays a key role in stabilizing the supersolid state.  
There are many interesting problems to be explored in this context. 
An imbalanced fermionic system with $N_\uparrow \neq N_\downarrow$ may be 
particularly interesting, since the so-called Fulde-Ferrell-Larkin-Ovchinnikov superfluid state with a periodically modulated order parameter might emerge and compete with the DW state, giving rise to a novel supersolid state. 

\acknowledgments
This work was partly supported by a Grant-in-Aid from the Ministry 
of Education, Culture, Sports, Science, and Technology of Japan 
[20740194 (A.K.), 20540390 (S.S.), 19014013 and 20029013 (N.K.)].
Some of the computations were performed at the Supercomputer Center at the 
Institute for Solid State Physics, University of Tokyo.

%

\end{document}